\documentclass[12pt]{article}
\usepackage{amssymb}
\def\beq{\begin{equation}}
\def\eeq{\end{equation}}
\usepackage[all,2cell,dvips]{xy}

\def\IR{\relax{\rm I\kern -.18em R}}

\begin{document}
\title{Supersymmetry breaking, conserved charges and stability in $N=1$ Super KdV }
\author{ \Large A. Restuccia*, A. Sotomayor**}
\maketitle{\centerline{*Departamento de F\'{\i}sica}}
\maketitle{\centerline{Universidad de Antofagasta}
\maketitle{\centerline{*Departamento de F\'{\i}sica}}
\maketitle{\centerline{Universidad Sim\'on Bol\'{\i}var }
\maketitle{\centerline {**Departamento de Matem\'{a}ticas,}}
\maketitle{\centerline{Universidad de Antofagasta}}
\maketitle{\centerline{e-mail: arestu@usb.ve, asotomayor@uantof.cl
}}
\begin{abstract}
We analyse the non-abelian algebra and the supersymmetric
cohomology associated to the local and non-local conserved
charges of N=1 SKdV under Poisson brackets. We then consider the
breaking of the supersymmetry and obtain an integrable model in
terms of Clifford algebra valued fields. We discuss the remaining
conserved charges of the new system and the stability of the
solitonic solutions.
\end{abstract}


Pacs: 12.60.Jv, 02.30.lk, 11.30.-j, 05.45. Yv, 02.30. Jr

\section{Introduction}Local and non-local conserved charges are
relevant in order to analyze perturbative and non-perturbative
aspects of a field theory. It is known that the local conserved
charges of $N=1$ SKdV are related to the superconformal algebra.
It is then interesting to look for an extension of the abelian
algebra, spannned by the local conserved charges of $N=1$ SKdV,
in terms of the non-local conserved charges. We obtain such
extension as a non-abelian algebra realized in terms of Poisson
brackets and discuss its relation to the supersymmetric structure
of $N=1$ SKdV. We then consider the susy breaking on the
symmetries associated to the higher dimensional conserved
charges. The resulting system, although has less symmetries than
the original SKdV one, has solitonic solutions with nice stability
properties.

\section{Non-abelian algebra of conserved charges and supersymmetric cohomology for $N=1$ SKdV}
The susy $N=1$ Korteweg-de Vries equation in superfield form is
given by \beq \Phi_t=D^6\Phi+3D^2\left(\Phi D\Phi\right), \eeq
where $\Phi=\xi+\theta u$ and $\xi,u$ are fields with values on
the odd and even parts respectively of a given Grassmann algebra
$\Lambda$ with $n$ generators and a distinguished one $\theta$,
and $D=\partial_\theta+\theta\partial_x$ is the covariant
derivative with the property $D^2=\partial_x$ acting on
superfields.

In components (1) is equivalent to
\beq\begin{array}{l}u_t=u^{\prime\prime\prime}+6uu^\prime-3\xi\xi^{\prime\prime}
\\ \xi_t=\xi^{\prime\prime\prime}+3{(\xi u)}^\prime.
\end{array}\eeq
The system (2) is integrable in the sense that it admits an
infinite number of local conserved charges. It is invariant under
the supersymmetric transformation defined by $\delta\Phi=\eta
Q\Phi$, where $\eta$ is an odd parameter and
$Q=-\partial_\theta+\theta\partial_x.$ It is also known to have an
infinite sequence of odd and even non-local conserved charges.
They can be obtained using simultaneously an auxiliary equation,
the so called super-Gardner equation \beq\chi_t=D^6\chi+3D^2(\chi
D\chi)-\varepsilon^23(D\chi)D^2(\chi D\chi)\eeq and the
super-Gardner transformation $\Phi=\chi+\varepsilon
D^2\chi-\varepsilon^2\chi D\chi$, relating (1) with (3) and
specifically solutions between them. The infinite sequence of
local conserved charges of the $N=1$ Super KdV equation were
obtained by Mathieu \cite{Mathieu1} from the one conserved charge
of the super-Gardner equation \beq G_1=\int
dxd\theta\chi=\sum_{n=0}\epsilon^{2n}H_{2n+1},\eeq using the
inverse Gardner transformation.
 The infinite sequence of fermionic non-local
conserved charges were obtained by Dargis and Mathieu
\cite{Dargis} and Kersten \cite{Kersten}. This infinite sequence
can be obtained also from one non-local conserved charge of the
super-Gardner equation \cite{Andrea1}  \beq
G_{\frac{1}{2}}^{NL}=\int dxd\theta \left(\frac{\exp(\epsilon
D^{-1}\chi)-1}{\epsilon}\right)=\sum_{n=0}\epsilon^nH_{n+\frac{1}{2}}^{NL}.\eeq
And finally, the infinite sequence of bosonic non-local conserved
charges was derived \cite{Andrea2} from the non-local conserved
charge for super-Gardner given by \beq H_G=\frac{1}{2}\int
dxd\theta D^{-1}\left\{D\left[\frac{\exp\left(\epsilon D^{-1}\chi
\right)-1}{\epsilon}\right] D^{-1}\left[\frac{\exp\left(-\epsilon
D^{-1}\chi\right)-1}{\epsilon}\right]\right\}=\sum_{n=0}\epsilon^{n}H_{n+1}^{NL}.\label{consgard2}\eeq
In (4),(5) and (6) the subscripts of the right members denote the
dimension of the respective conserved charges.

It is interesting to notice that the Gardner deformation for
$N=2$ supersymmetric $a=4$ KdV equation has also been recently
solved \cite{Kiselev1,Kiselev2}.

Given a conserved charge for equation (1) of the form  $H=\int
dxd\theta h,h\in C_I^\infty$, where $C_I^\infty$ is
 the ring of integrable superfields, we define
\beq\delta_QH:=\int dxd\theta Qh,\eeq where $\delta_Q$ satisfies
$\delta_Q \delta_Q=0$. (7) defines a cohomology on the conserved
charges of super-KdV \cite{Andrea2}. In fact, for example we have
\beq\delta_QH_{2n+1}=0,n=0,1,\ldots,\label{cohom0}\eeq hence
$H_{2n+1}$ must be $\delta_Q$ of some other conserved charge, in
particular for $n=1$ we have \beq \delta_QH_\frac{1}{2}^{NL}=H_1.
\label{cohom4}\eeq In the same way we obtain a relation between
the non-local odd and even conserved charges, for example
\beq\delta_QH_1^{NL}=H_{\frac{3}{2}}^{NL}-\frac{1}{2}H_1H_{\frac{1}{2}}^{NL}.\eeq
The general formula is
\beq\sum_{n=0}\epsilon^{n}\delta_QH_{n+1}^{NL}=\sum_{n=0}\epsilon^{2n}H_{2n+\frac{3}{2}}^{NL}-\frac{1}{2\epsilon}
\left[\exp\left(\sum_{n=0}\epsilon^{2n+1}H_{2n+1}\right)-1\right]
\left[\sum_{n=0}{(-\epsilon)}^{n}H_{n+\frac{1}{2}}^{NL}\right].\label{genform}\eeq
We then consider the Poisson bracket of superfields $\Phi$ at two
 different points in superspace as given in \cite{Mathieu1}:
 \beq
\left\{\Phi(x_1,\theta_1),\Phi(x_2,\theta_2)\right\}=P_1\Delta,\label{brac1}
\eeq  where
$P_1=D_1^5+3\Phi_1D_1^2+(D_1\Phi_1)D_1+2(D_1^2\Phi_1)$ and
$\Delta=\delta(x_1-x_2)(\theta_1-\theta_2).$ With this bracket,
using the supersymmetric cohomology and after some calculations,
it is possible to derive the Poisson algebra of local and
non-local conserved charges, in particular we obtain the following
relations:
\begin{eqnarray}& & \left\{H_{\frac{1}{2}}^{NL},H_{\frac{1}{2}}^{NL}\right\}=H_1,   \\
& &
\left\{H_{\frac{1}{2}}^{NL},H_{\frac{3}{2}}^{NL}\right\}=\frac{1}{2}H_1^2,
\\ & &
\left\{H_{\frac{1}{2}}^{NL},H_1^{NL}\right\}=-\frac{1}{2}H_1H_{\frac{1}{2}}^{NL}+H_{\frac{3}{2}}^{NL},
\end{eqnarray} and
\beq\left\{H_{\frac{3}{2}}^{NL},H_{\frac{3}{2}}^{NL}\right\}=-H_3+\frac{1}{3}{(H_1)}^3.
\eeq This gives the first steps in order to construct the complete
non-abelian algebra of conserved charges for $N=1$ super-KdV
equation. This problem is under development.

The local and non-local conserved charges are related to
symmetries of the $N=1$ SKdV equation. We are now interested in
considering the breaking of supersymmetry which will induce a
breaking of the symmetries associated with the conserved charges.

 \section{Supersymmetry breaking of SKdV and stability of the ground state and the one-soliton solutions }In order to
 break the supersymmetry we consider the system \cite{Adrian}
 \beq\begin{array}{l}u_t=-u^{\prime\prime\prime}-uu^\prime-\frac{1}{4}{\left(\mathcal{P}\left(\xi\bar{\xi}\right)\right)}^\prime
\\ \xi_t=-\xi^{\prime\prime\prime}-\frac{1}{2}{(\xi u)}^\prime,
\end{array}\label{Clifford}\eeq where the fields $u$ and $\xi$
takes values on a Clifford algebra instead of being Grassmann
algebra valued. In order to compare with the stability analysis
of KdV in Benjamin \cite{Benjamin} we have redefined the $u$ and
$\xi$ fields in (2) with convenient factors. We thus take $u$ to
be a real valued field while $\xi$ to be an expansion in terms of
the generators $e_i,i=1,\ldots$ of the Clifford algebra:
\beq\xi=\sum_i
\varphi_ie_i+\sum_{ij}\varphi_{ij}e_ie_j+\sum_{ijk}\varphi_{ijk}e_ie_je_k+\cdots\eeq
where \beq e_ie_j+e_je_i=-2\delta_{ij}\eeq and
$\varphi_i,\varphi_{ij},\varphi_{ijk},\ldots$ are real valued
functions. We define $ \bar{\xi}=\sum_{i=1}^\infty
\varphi_i\bar{e_i}+\sum_{ij}\varphi_{ij}\bar{e_j}\bar{e_i}+\sum_{ijk}\varphi_{ijk}\bar{e_k}\bar{e_j}\bar{e_i}+\cdots$
where $\bar{e_i}=-e_i$. We denote as in superfield notation the
body of the expansion those terms associated with the identity
generator and the soul the remaining ones. Consequently the body
of $ \xi\bar{\xi}$, denoted by $ \mathcal{P}(\xi\bar{\xi})$, is
equal to
$\Sigma_i\varphi_i^2+\Sigma_{ij}\varphi_{ij}^2+\Sigma_{ijk}\varphi_{ijk}^2+\cdots$
In what follows, without loss of generality, we rewrite
$\mathcal{P}(\xi\bar{\xi})=\Sigma_i
\varphi_i^2+\Sigma_{ij}\varphi_{ij}^2+\Sigma_{ijk}\varphi_{ijk}^2+\cdots$
simply as $\mathcal{P}(\xi\bar{\xi})=\Sigma_i\varphi_i^2$. The
system (\ref{Clifford}) is not invariant under supersymmetric
transformations and has not a conserved charge of dimension 7,
that is there is no analogue of $H_7$ in SKdV or KdV systems  but
has the following conserved charges:
\beq\begin{array}{l}\hat{H}_{\frac{1}{2}}=\int_{-\infty}^{+\infty}\xi
dx,\hspace{3mm}\hat{H}_1=\int_{-\infty}^{+\infty} u dx
\\ V\equiv \hat{H}_3=\frac{1}{2}\int_{-\infty}^{+\infty}\left(u^2+\mathcal{P}\left(\xi\bar{\xi}\right)
\right)dx,\hspace{3mm} \\ M\equiv
\hat{H}_5=\frac{1}{2}\int_{-\infty}^{+\infty}
\left(-\frac{1}{3}u^3-\frac{1}{2}u\mathcal{P}\left(\xi\bar{\xi}\right)+{(u^\prime)}^2+\mathcal{P}\left(\xi^\prime\bar{\xi}^\prime\right)\right)dx.
\end{array}\eeq
It is interesting to remark that the following non-local conserved
charge of Super KdV \cite{Andrea2} is also a non-local conserved
charge for the system (17), in terms of the Clifford algebra
valued field $\xi$,
\[\int_{-\infty}^\infty \xi(x)\int_{-\infty}^x\xi(s)dsdx.\]
 However the non-local conserved charges of Super KdV in \cite{Dargis} are not conserved by the system (17). For example, \[\int_{-\infty}^\infty
u(x)\int_{-\infty}^x\xi(s)dsdx\] is not conserved by (17).

We now consider the stability of solutions for the system (17) in
the sense of Liapunov \cite{Adrian}. In particular we take the
same definition as in \cite{Benjamin}: $( \hat{u},\hat{\xi})$, a
solution of (17), is stable if given $\epsilon$ there exists
$\delta$ such that for any solution $(u,\xi)$ of (17), satisfying
at $t=0$

\beq
d_I\left[\left(u,\xi\right),\left(\hat{u},\hat{\xi}\right)\right]<\delta
\eeq

then

\beq
d_{II}\left[\left(u,\xi\right),\left(\hat{u},\hat{\xi}\right)\right]<\epsilon\eeq

for all $t\geq 0$.

$d_I$ and $d_{II}$ denote two distances to be defined.

We denote by $\|\hspace{2mm} \|_{H_1}$ the Sobolev norm

\beq {\|(u,\xi)
\|_{H_1}}^2=\int_{-\infty}^{+\infty}\left[\left(u^2+\Sigma_{i=0}^\infty
\varphi_i^2 \right)+\left ({u^\prime}^2+\Sigma_{i=0}^\infty
{\varphi_i^\prime}^2\right)\right]dx. \eeq It can be proved that
\beq {\|(u,\xi) \|_{H_1}}^2\leq V+M+\frac{1}{\sqrt{2}}V{\|(u,\xi)
\|_{H_1}}\label{prioribound}, \eeq and this gives a priori bound
for solutions to the system (17). The stability of the ground
state solution $ \hat{u}=0,\hat{\xi}=0$ is a direct consequence of
(\ref{prioribound}) taking $d_I$ and $d_{II}$ to be the Sobolev
norm $\|(u-\hat{u},\xi-\hat{\xi}) \|_{H_1}.$

The system (17) has also solitonic solutions, for example:
$u(x,t)\equiv \phi(x,t)=3\mathcal{C}\frac{1}{\cosh^2(z)}, z\equiv
\frac{1}{2}\mathcal{C}^{\frac{1}{2}}(x-\mathcal{C}t),\xi(x,t)=0.$
$\phi(x,t)$ is the one-soliton solution of KdV equation. We now
consider the stability of the one-soliton solution $
\hat{u}=\phi,\hat{\xi}=0.$ The proof of stability is based on
estimates for the $u$ field which are analogous to the one
presented in \cite{Benjamin,Bona} while a new argument will be
given for the $\xi$ field. The distances we will use are \beq
d_I\left[\left(u_1,\xi_1\right),\left(u_2,\xi_2\right)\right]=\|\left(u_1-u_2,\xi_1-\xi_2\right)\|_{H_1}
\eeq \beq
d_{II}\left[\left(u_1,\xi_1\right),\left(u_2,\xi_2\right)\right]=\inf_\tau\|\left(\tau
u_1-u_2,\xi_1-\xi_2\right)\|_{H_1} \eeq where $\tau u_1$ denotes
the group of translations along the $x$-axis. $d_{II}$ is a
distance on a metric space obtained by identifying the
translations of each $u\in H_1(\mathbb{R})$. $d_{II}$ is related
to a stability in the sense that a solution $u$ remains close to
$ \hat{u}=\phi$ only in shape but not necessarily in position.

We first assume that

\beq V(u,\xi)=V( \hat{u},\hat{\xi})=V(\phi,0) \eeq

and

\beq \int_{-\infty}^{+\infty}\xi dx=
\int_{-\infty}^{+\infty}\hat{\xi}dx=0 \eeq and at the end of the
argument we relax this conditions to get the most general
formulation of the stability problem. It is then possible to
prove, after long reasoning that \beq\Delta M\equiv
M(u,\xi)-M(\phi,0)\leq\left[\max\left(1,\mathcal{C}\right)+\frac{1}{3\sqrt{2}}\delta\right]{\|\left(h,\xi\right)\|}_{H_1}^2\eeq
and that \beq\Delta M\geq\frac{1}{6}l{\left\{
d_{II}\left[(u,\xi),(\phi,0)\right] \right\}}^2, \eeq and these
two inequalities, using that $\Delta M$ is a conserved quantity,
are sufficient to prove the stability of the one-soliton solution
for the system (17). To relax the conditions (27) and (28) we use
essentially the triangle inequality.

\section{Conclusions}We discussed the algebraic structure of the
local and non-local conserved charges of $N=1$ SKdV. We presented
a rigorous construction of the Poisson brackets of these
non-local conserved charges, this is in general an open problem
in field theory.

The non-abelian algebra has been worked out partially, we hope to
provide the complete algebraic structure in a future work. Besides
we considered the supersymmetry breaking of $N=1$ SKdV, this is
always an interesting problem to study. The resulting system is
integrable, in the sense that it has solitonic solutions, but, of
course, it is not supersymmetric. Moreover the infinite sequence
of local and non-local conserved charges becomes certainly
reduced, possibly only to a finite set. Consequently the
associated symmetries have also been broken. In spite of the fact
that the new integrable system has less symmetries, its solitonic
solutions are stable in the Liapunov sense. The problem of
stability of the solutions of the supersymmetric KdV system is
also an open problem. Our result is a contribution in this sense.

 $\bigskip$

\textbf{Acknowledgments}

A. R. and A. S. are partially supported by Project Fondecyt
1121103, Chile.


\begin{thebibliography}{}
\bibitem{Mathieu1}{P. Mathieu, J. Math. Phys. 29, 2499
(1988).}
\bibitem{Dargis}{P. Dargis and P. Mathieu, Phys. Lett. A 176,
67-74 (1993).}
\bibitem{Kersten}{P. H. M. Kersten,  Phys. Lett. A 134, 25 (1988).}
\bibitem{Andrea1}{S. Andrea, A. Restuccia and A. Sotomayor, J. Math.
Phys. 46, 103517 (2005).}
\bibitem{Andrea2}{S. Andrea, A. Restuccia and A. Sotomayor, Phys. Lett.
A 376, 245-251 (2012).}
\bibitem{Kiselev1}{V.Hussin, A. V. Kiselev, A. O. Krutov, and T.
Wolf, J. Math. Phys. 51(8), 083507 (2010).}
\bibitem{Kiselev2}{A. V. Kiselev and O. Krutov, J. Math. Phys. 53, 103511 (2012).}
\bibitem{Adrian}{A. Restuccia and A. Sotomayor, arXiv: 1207.5870v1[math-ph] (2012).}
\bibitem{Benjamin}{T. B. Benjamin, Proc. R. Soc. Lond. A. 328,
153-183 (1972).}
\bibitem{Bona}{J. Bona, Proc. R. Soc. Lond. A.  344, 363-374
(1975).}


\end{thebibliography}
\end{document}